\documentclass[10pt,letterpaper]{article}
\usepackage[top=2.5cm,bottom=2.5cm,left=3cm,right=3cm]{geometry}

\usepackage{amsmath,amssymb}
\usepackage{txfonts}

\usepackage{changepage}

\usepackage[utf8x]{inputenc}

\usepackage{textcomp,marvosym}

\usepackage{cite}

\usepackage{nameref,hyperref}

\usepackage[right]{lineno}

\usepackage{microtype}
\DisableLigatures[f]{encoding = *, family = * }

\usepackage[table]{xcolor}

\usepackage{array}

\newcolumntype{+}{!{\vrule width 2pt}}

\newlength\savedwidth

\usepackage[aboveskip=1pt,labelfont=bf,labelsep=period,justification=raggedright,singlelinecheck=off]{caption}

\makeatletter
\renewcommand{\@biblabel}[1]{\quad#1.}
\makeatother

\usepackage{lastpage,fancyhdr,graphicx}
\fancyheadoffset[L]{2.25in}
\fancyfootoffset[L]{2.25in}
\usepackage{xspace}
\usepackage{cleveref}
\usepackage{bm}
\usepackage{booktabs}

\newcommand{\refeq}[1]{\cref{eq:#1}}

\newcommand{\Fig}[1]{Fig~\ref{fig:#1}}

\newcommand{\bigO}[1]{\ensuremath{\mathcal{O}(#1)}}

\newcommand{\tmin}{\ensuremath{t_{min}}\xspace}

\newcommand{\kappad}{\ensuremath{\kappa_d}\xspace}

\newcommand{\taud}{\ensuremath{\tau_d}\xspace}
\newcommand{\npat}{43\xspace}
\newcommand{\params}{\ensuremath{R_0,v,k,\tau}}
\newcommand{\paramshat}{\ensuremath{\hat{R_0},\hat{v},\hat{k},\hat{\tau}}}
\newcommand{\tmes}{\ensuremath{t_\textrm{mes}}}

\begin{document}
\vspace*{0.2in}

\begin{flushleft}
{\huge
\textbf{Predicting regrowth of low-grade gliomas after radiotherapy} 
}
\newline
\\
St\'ephane Plaszczynski \textsuperscript{1,2*},
Basile Grammaticos \textsuperscript{1,2},
Johan Pallud \textsuperscript{3, 4, 5},
Jean-Eric Campagne \textsuperscript{1,2},
and Mathilde Badoual \textsuperscript{1,2}
\\
\bigskip
\textbf{1} Universit\'e Paris-Saclay, CNRS/IN2P3, IJCLab, 91405 Orsay, France;
\\
\textbf{2} Universit\'e Paris-Cit\'e, IJCLab, 91405 Orsay, France;
\\
\textbf{3} Department of Neurosurgery, GHU Paris 
Sainte-Anne Hospital, 75014 Paris, France;
\\
\textbf{4} Universit\'e  de Paris, Sorbonne Paris Cité, Paris, France;
\\
\textbf{5} Inserm, U1266, IMA-Brain, Institut de Psychiatrie et
Neurosciences de Paris, France.

\bigskip

%
%
*~ stephane.plaszczynski@ijclab.in2p3.fr

\end{flushleft}

\section*{Abstract}
Diffuse low grade gliomas are invasive and incurable brain tumours
that inevitably transform into higher grade ones. A classical
treatment to delay this transition is radiotherapy (RT).
Following RT, the tumour gradually shrinks during a period of
typically 6 months to 4 years before regrowing.
To improve the patient's health-related quality of life and help
clinicians build personalised follow-ups, one would
benefit from predictions of the time during which the tumour is expected to
decrease. The challenge is to provide a reliable estimate of this
regrowth time shortly after RT (i.e. with few data), although patients react differently to
the treatment.
To this end, we analyse the tumour size dynamics from a batch of 20
high-quality longitudinal data, and propose a simple and robust
analytical model, with just 4 parameters. From the study of their
correlations, we build a statistical constraint that helps determine the
regrowth time even for patients for which we have only a few
measurements of the tumour size.
We validate the procedure on the data and predict the regrowth
time at the moment of the first MRI after RT,
with  precision of, typically, 6 months.
Using virtual patients, we study whether some forecast is still
possible just three months after RT.
We obtain some reliable estimates of the regrowth time in 75\% of the
cases, in particular for all ``fast-responders''. The remaining 25\%
represent cases where the actual regrowth
time is large and can be safely estimated with another measurement a year
later. These results show the feasibility of making personalised
predictions of the tumour regrowth time shortly after RT.

\section*{Introduction}

Diffuse gliomas are primary brain tumours originating from glial
cells (oligodendrocytes and/or astrocytomas). In its 2016 classification, The World Health Organization defines four grades \cite{who16}: while the first grade gliomas are benign,
second grade gliomas (or low grade gliomas, LGG) are invasive, 
growing at a rate of $2$ to $8$ mm/year \cite{mandonnet03} in
diameter, but without involving metastasis or necrosis. Unfortunately, they
cannot be cured by oncological treatments \cite{pallud10bis}
so one needs to contain their growth as long as possible, before they transform into grade III
and IV (glioblastomas) with a dramatically low survival rate. 
LGG are detected with magnetic
resonance imaging (MRI) scans under a T2-FLAIR sequence. Since they are diffuse tumours that extend beyond the observed boundaries
\cite{kelly87, pallud10}, the uncertainty on their size is irreducible.
Classical treatments include resection (when possible), chemotherapy and
radiotherapy (RT) \cite{Soffietti:2010}.
 
Standard conformational radiotherapy for LGG is generally performed
during 6 weeks (5 days a week) and the classical dose is around 50 Gy.
Irradiation of gliomas involves a large number of 
physical processes \cite{wang18} and its effect varies across patients. However, some general
features emerge: the tumour shrinks during a period that
varies between a few months and several years, before regrowing at a rate
similar to the one observed before radiotherapy.

Mathematical modelling of natural and under treatment tumour growth has a long and rich history (in particular for gliomas, one can refer to the recent review \cite{altrock15}). 
For invasive tumour such as gliomas that cannot be removed by surgery,
one aspect that is of special interest for clinicians is the response
of tumour to treatments and in particular, radiotherapy
\cite{rockne10,badoual14,perez15,boudia19,Adenis:2021}. 
Its primary goal is to optimize treatments ``virtually'': for example,
choosing the optimal radiation fractionation of doses
\cite{galochkina15}, finding the best way to combine it to
chemotherapy \cite{ayala2021} or studying its interplay with the
immune system \cite{bekker22}. 
Beyond describing qualitatively the different processes at stake, the
real usefulness of a model would be to predict the response of
individual patients to a treatment, even before the end of the
treatment. Such predictions would allow the clinician to personalise
the follow-up (and the treatment) for each patient. 
There has been some attempts to predict tumour growth and the effects
of treatments on individual patients. If purely statistical or
image-based models can be used to predict glioma growth
\cite{elazab21}, mechanistic models are usually used for instance 
to predict the metastatic relapse in
breast cancer \cite{nicolo20}, tumour growth in leukaemia and ovarian
cancer \cite{mascheroni21}, response of high grade gliomas to
chemoradiation \cite{hormuth21}, or the patient-specific evolution of
resistance in the context of prostate cancer \cite{brady-nicholls20}.

For low-grade gliomas, individualized predictions from the tumour size
dynamics and genetic characteristics, have been made for
the response to a chemotherapy treatment \cite{Mazzocco:2015}. To our knowledge, such individual predictions do
not exist in the case of low-grade glioma and RT.  
In this article, we show that it is possible to predict the evolution
of LGGs under RT, for individual patients, with an approach based on a
practical mechanistic model, even in the case where the number of
patients is not sufficient to apply standard machine-learning techniques.

In order to be used for predictions, a model should have a limited number of parameters. Models that are too detailed are useful to describe qualitatively the tumour evolution but usually involve too many unknown parameters \cite{brady19}. Given the scarcity of clinical data, only a small number of parameters can be deduced. The goal here is to keep as
few as possible parameters  but still to capture the essential dynamics of
the tumour.

In a previous work \cite{Adenis:2021}, we analysed a large number (\npat) of
LGG radial evolutions under RT
and proposed a physically motivated model, with 4 parameters, that fitted
well all the profiles of tumour evolution during patient's follow-up. 
Following that work, we now try to make predictions using that model.
This is challenging
given the variety of possible in-vivo reactions to radiation.
We have chosen to focus on the moment when the tumour stops shrinking and
starts to regrow, what we call in the following the ``regrowth time''. 
This is an essential feature of the tumour dynamics for two reasons.
First, the patients often ask their clinician when the tumour will regrow 
in order to plan some major life projects (as having a child,
travelling, retiring, etc.). This would be a valuable information to improve their
life-quality.
Second, the the dates for the next MRIs are currently fixed and not optimal on an individual basis. By making
predictions we may adjust them  more precisely for personalised follow-ups.

\section{Materials and Methods}
\label{sec:method}

\subsection{Standard protocol approvals, registrations and patient consents} 
The study received required authorizations (IRB\#1: 2021/20) from the
human research institutional review board (IRB00011687). The
requirement to obtain informed consent was waived according to French
legislation (observational retrospective study).

\subsection{The patients} 
\label{sec:patients}

We had at our disposal a set of \npat patients with LGGs who were diagnosed at
the Sainte-Anne Hospital (Paris, France) from 1989 to 2000. These
patients were selected according to precise criteria that are detailed
elsewhere~\cite{pallud12a}. In~short, only adults with typical LGGs
(that is, no angiogenesis and, thus, no contrast enhancement on
gadolinium-T1 images), available clinical and imaging follow-ups
before, during, and~after RT, and RT as their first oncological
treatment except for stereotactic biopsies were eligible. The external
conformational RT was given using the same methodology (total dose,
50.4--54 Gy; \mbox{6-week period}) at 2 outside institutions.
The patients had an MRI follow-up before, during,~and after RT. Three
tumour diameters in the axial, coronal, and sagittal planes on each MRI
image with T2-weighted and FLAIR sequences were measured manually. The~mean radiological tumour radius was defined as half the geometric mean
of these three diameters and was measured as a function of time. The 
error bars for the measured mean radii were estimated by
clinicians and were set to $\pm 1$ mm. From~this cohort, we discarded 
the patients that did not have any sign of tumour regrowth at the last
time point or~those that had fewer than five time points in
their~follow-up. 

\subsection{The model}
\label{sec:model}

A biologically motivated model with the effect of RT on LGG has been 
presented in \cite{Adenis:2021} and validated by the fits on \npat
patient follow-ups. It is based on a
standard diffusion-proliferation equation \cite{cruywagen95} and RT is modelled with
a time-dependent death rate ($\kappa_D(t)$). The evolution of the glioma
cell density then follows the equation
\begin{align}
  \label{eq:oldmodel}
  \dfrac{\partial \rho}{\partial t}=D\Delta\rho+[\kappa-\kappa_D(t)] \rho
  (1-\rho),
\end{align}
where $\rho(r,t)$ is a function of the radius $r$ (assuming a spherical
symmetry) and time $t$
(conventionally set to zero at the beginning of RT), $D$ is the
diffusion coefficient and $\kappa$ the proliferation rate.
In its most simple (thus predictive) form, the death-term is characterized by an
amplitude and a characteristic time
\begin{align}
 \kappa_D(t)=\kappad e^{-t/\taud} \quad \textrm{for}~ t\geq0,
 \end{align}
and is considered as null before RT.

Assuming that the tumour growth-rate when patients consult is already in the
asymptotic state, i.e. that it evolves linearly with a speed
$v=\sqrt{2D\kappa}$, and neglecting diffusion after RT, the radius
evolution can be
approximated by \cite{Adenis:2021}

\begin{align}
  \label{eq:rast}
  R(t)=R_0 +vt - v \taud \dfrac{\kappad}{\kappa}(1-e^{-t/\taud}).
\end{align}

\begin{figure}[!ht]
  \centering
  \includegraphics[width=0.7\textwidth]{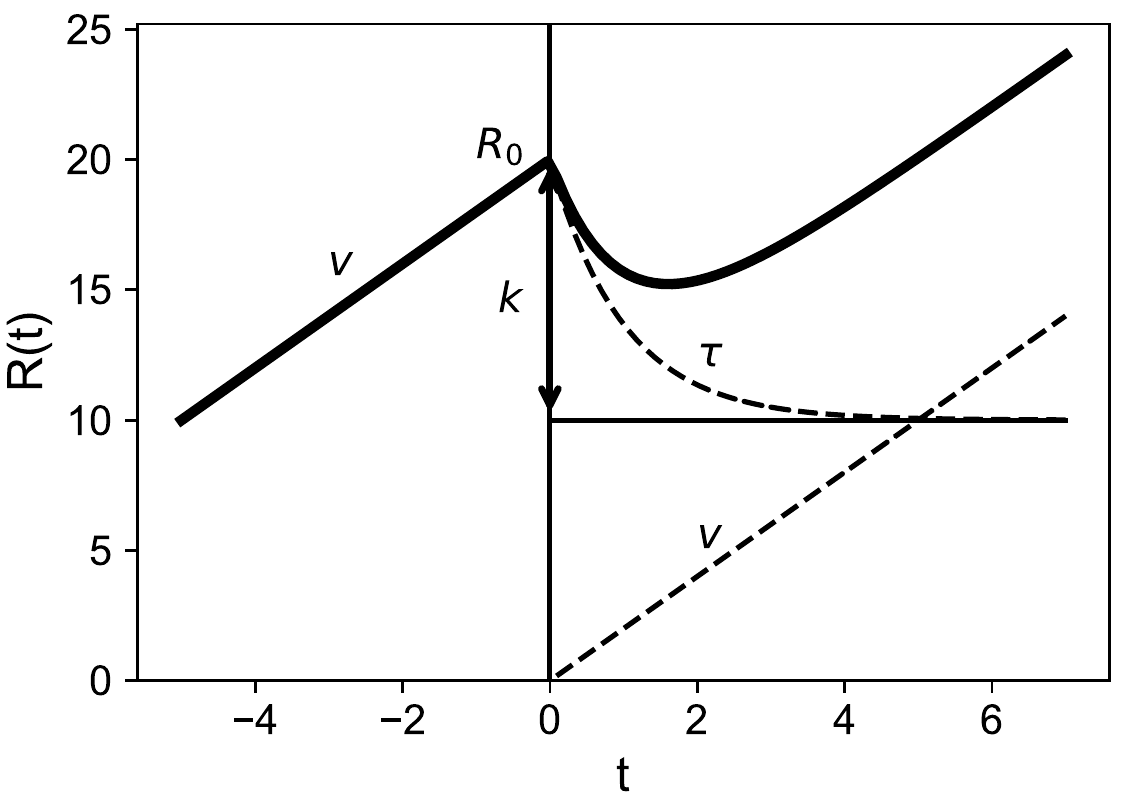}
  \caption{ \textbf{Illustration of the analytical model describing
    the tumour radial evolution}. Before RT ($t<0$),
    the radius evolves linearly
    with the asymptotic speed $v$ and reaches $R_0$ at $t=0$. Then it
    becomes the sum of an exponential decay of amplitude $R_0-k$ and
    characteristic time $\tau$ (death term) and a linear $vt$
    regrowth.}
  \label{fig:fig1}
\end{figure}

Inspired by this formula, we simplify the model \refeq{oldmodel} by
proposing a purely geometrical one in the form
\begin{align}
  \label{eq:model}
  R(t)=R_0 +vt - k(1-e^{-t/\tau}).
\end{align}
which has 4 free parameters: \params. We emphasize that \refeq{model}
should be considered as an \textit{ad-hoc} model that cannot be related to
the one obtained by solving numerically \refeq{oldmodel} since 
\refeq{rast} neglects diffusion. This simple geometrical model has the considerable
advantage of being analytical. The role of each terms
is clear and sketched in \Fig{fig1}. It captures the 3 phases of
the evolution: first the linear growth, then the exponential decay of
a fraction of the tumour and therefore of its radius, third, the
regrowth with the same velocity as before RT.

\begin{figure}	
\includegraphics[width=\textwidth]{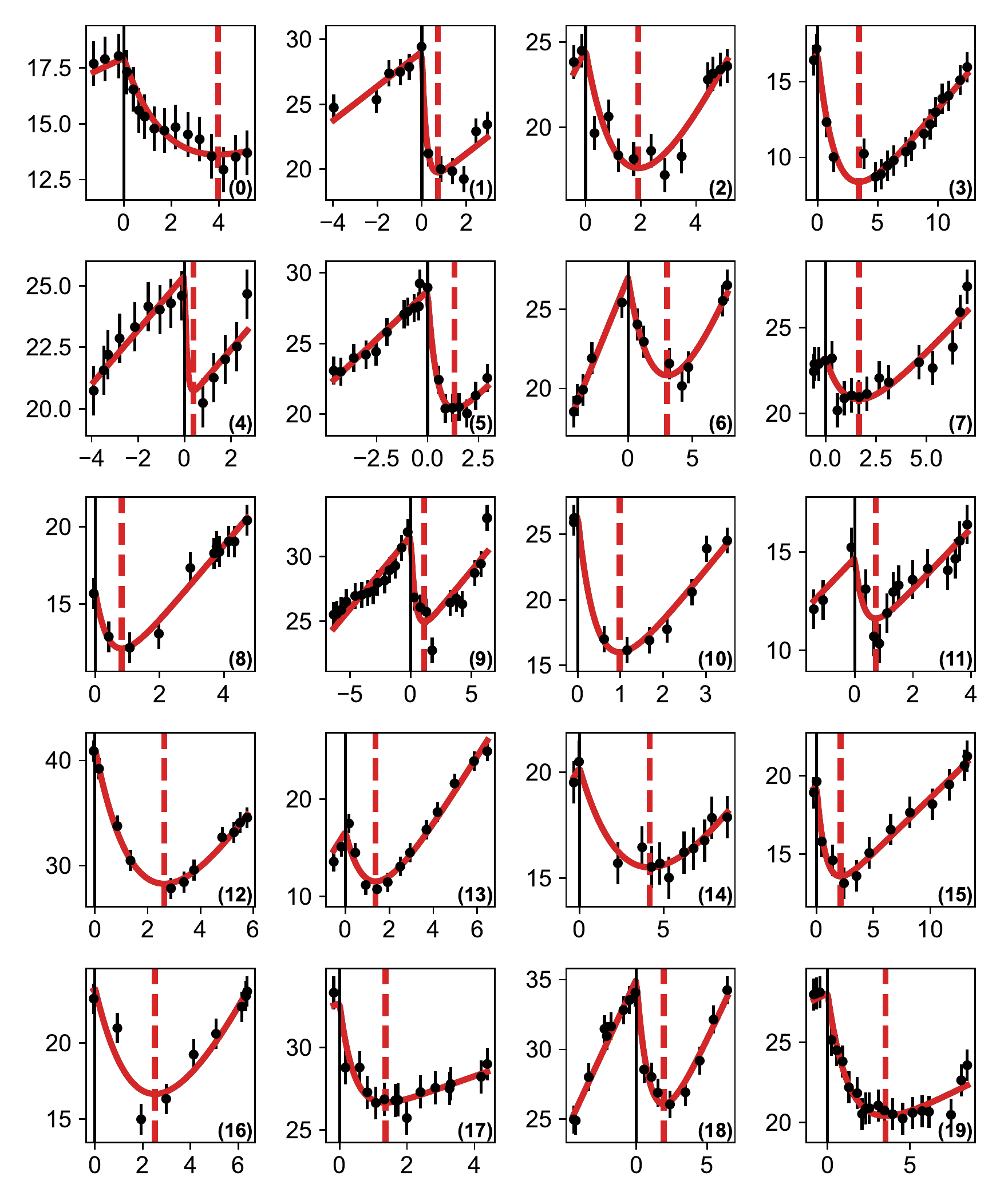}
\caption{\textbf{Comparison between the measured values of the tumour radius
  and the bestfit model for 
  20 patients}. The points represent the measured values and the red line our
model obtained by minimizing \refeq{chi2}. 
The abscissa represent time in years (with the origin
  set at RT)
  and the ordinate the tumour radius (in mm). The error bars on the measurements
  are of 1 mm. The dashed vertical red line shows the model minimum, i.e. the
  moment regrowth starts.
    \label{fig:bf}}
\end{figure}

To test whether this model does fit our data appropriately, we
construct a classical objective function as the mean squared error from the set of measured
values $\{t_i,R_i\}$
\begin{align}
  \label{eq:chi2}
  \chi^2(\params)=\sum_{i=1}^N [R_i-R(t_i;\params)]^2/\sigma_i^2
\end{align}
where $\sigma_i=1$ mm. This is a 4-parameter real valued
function that we minimize easily with a standard optimization
algorithm \cite{Byrd:1995} since the model is analytical. To obtain
physical results, we impose as limits that all parameters be positive and that the
radial asymptotic speed lie in the range $0.5\leq v \leq 4$ mm/yr \cite{mandonnet03}.
We then obtain the best-fit parameter values given in Table \ref{tab:bf}
and show, in \Fig{bf}, the comparison between the data and the fitted
model on a set of 20 patients who possess at least 9 data points.
The agreement is excellent. Although the results are quite similar to the ones obtained in
\cite{Adenis:2021}, we have considerably simplified the model 
and reduced drastically the run-time which will be useful later in making predictions.

\begin{table}
  \centering
\begin{tabular}{lrrrrr}
\toprule
id &    $R_0$ &    $v$ &     $k$ &  $\tau$ & \tmin\\
\midrule
(0)  & 17.95 & 0.50 &  7.29 & 1.98 & 3.95 \\
(1)  & 29.03 & 1.33 & 10.49 & 0.19 & 0.72 \\
(2)  & 24.56 & 3.24 & 17.69 & 1.40 & 1.91 \\
(3)  & 16.86 & 0.96 & 13.28 & 1.57 & 3.42 \\
(4)  & 25.40 & 1.11 &  5.20 & 0.10 & 0.39 \\
(5)  & 28.72 & 1.37 & 10.78 & 0.47 & 1.33 \\
(6)  & 27.00 & 1.99 & 17.00 & 2.39 & 3.04 \\
(7)  & 23.26 & 1.23 &  5.92 & 1.18 & 1.66 \\
(8)  & 15.83 & 2.45 &  6.84 & 0.45 & 0.82 \\
(9)  & 31.60 & 1.13 &  8.35 & 0.38 & 1.12 \\
(10) & 26.45 & 4.00 & 16.15 & 0.45 & 0.98 \\
(11) & 14.67 & 1.54 &  4.66 & 0.32 & 0.72 \\
(12) & 41.20 & 4.00 & 30.74 & 1.83 & 2.63 \\
(13) & 16.64 & 3.59 & 13.83 & 1.06 & 1.37 \\
(14) & 20.21 & 1.30 & 14.84 & 3.64 & 4.16 \\
(15) & 19.33 & 0.70 &  7.80 & 0.81 & 2.11 \\
(16) & 23.61 & 3.37 & 23.15 & 2.28 & 2.51 \\
(17) & 32.68 & 0.72 &  7.30 & 0.43 & 1.36 \\
(18) & 35.02 & 2.19 & 15.32 & 0.99 & 1.93 \\
(19) & 28.06 & 0.52 & 10.16 & 1.29 & 3.52 \\
\bottomrule
\end{tabular}
  \caption{\textbf{Parameters of the least-square solutions of
      \refeq{chi2} corresponding to the fits shown on \Fig{bf}}. The first columns represents the
    patients' ID, (\params) are the estimated parameters of
    the model, and
    the regrowth time (\tmin) is derived from them. 
    Lengths ($R_0,k$) are expressed in mm and times ($\tau,\tmin$) in years.}
  \label{tab:bf}
\end{table}

\subsection{Constraining the parameters space}
\label{sec:cons}

\begin{figure}[!ht]
  \centering
  \includegraphics[width=0.7\textwidth]{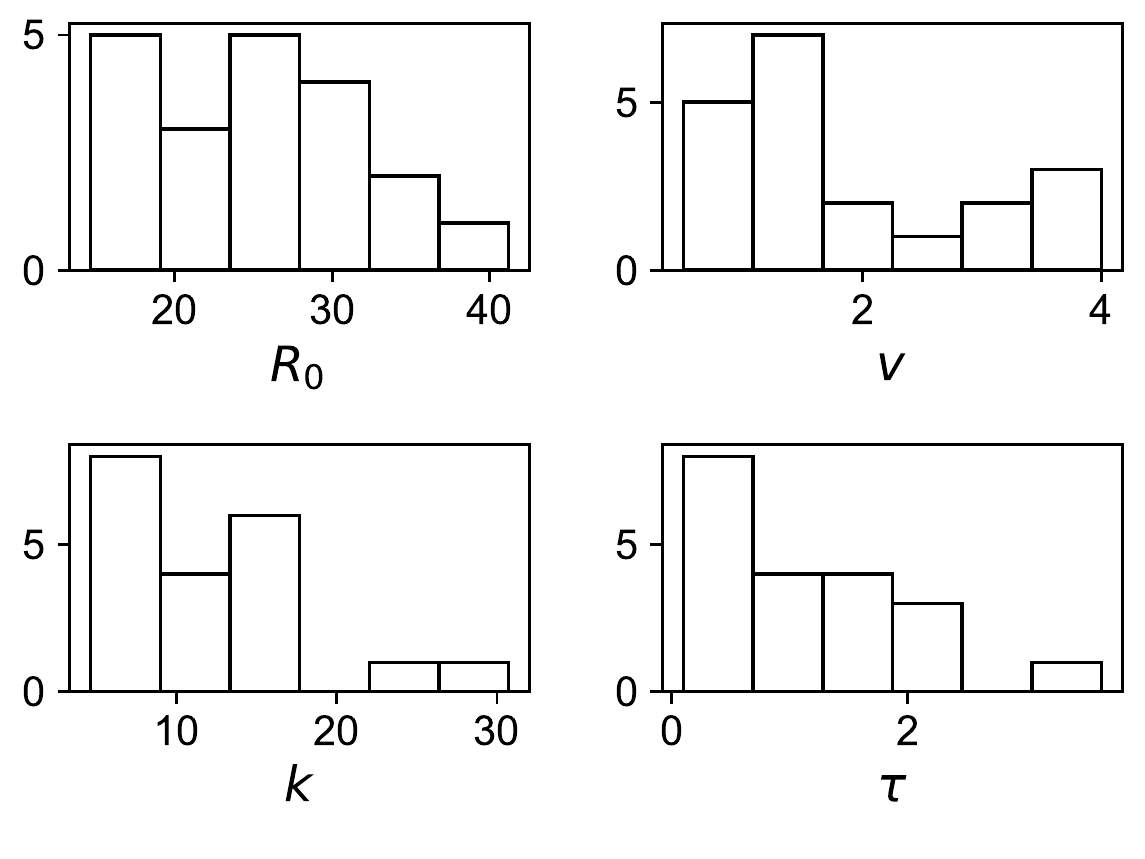}
  \caption{\textbf{Histograms of the bestfit parameters}}
  \label{fig:params}
\end{figure}

We now study whether some common features appear in our best-fit
parameters. \Fig{params} shows the histograms for each parameter on the 20 patients.

No parameter displays a clearly peaked distribution. For a given
patient, the expected parameters are random variables and a priori
unpredictable, although within some bounds.

We now consider the correlation between the variables by computing their
Pearson coefficients and show the results in Table \ref{tab:corr}. 

\begin{table}[!htt]
  \centering
\begin{tabular}{l|rrrr}
{} &    $R_0$ &   $v$ &   $k$ &   $\tau$ \\
\midrule
$R_0$ &  1 & 0.17 & 0.46 & -0.09 \\
$v$ &   & 1 & 0.73 &  0.10 \\
$k$ &   &  & 1 &  0.52 \\
$\tau$ &  &  &  &  1 \\
\end{tabular}
  \caption{\textbf{Correlation coefficients measured between the 20 bestfit
    parameters of our model}. Since the matrix is symmetric with ones on the
    diagonal we only show its upper half.}
  \label{tab:corr}
\end{table}
The structure is far from being diagonal, indicating non-trivial
correlations among most pairs of variables. Of particular interest is
the large ($k,v$) correlation since it relates a quantity defined before RT
($v$) to a one after RT ($k$).

To make use of the information in the most efficient way, we first decorrelate
the variables. This is performed by diagonalizing the covariance
matrix \footnote{which is always possible since the covariance matrix is by construction always positive-definite.}.
From the eigenvectors, we build the transformation matrix $\bm{T}$ that
projects our parameters $\bm{p}^T=(\params)$ onto an orthogonal basis where the new
variables $\bm{X}^T=(x_1,x_2,x_3,x_4)$ are uncorrelated.
From our data we measure the following projection matrix:

\begin{align}
  \bm{T}=&
           \begin{pmatrix}
             0.75 & 0.07 & 0.65&  0.02 \\
             0.65 & -0.13& -0.74& -0.11\\
             -0.06& -0.64&  0.17& -0.74\\
             0.01& -0.75&  0.05&  0.66
           \end{pmatrix}
\end{align}
and the linear change of variables is then simply
\begin{align}
\label{eq:prod}
X=\bm{T}p  
\end{align}
Considering the important terms in the matrix, we see that the first 2
lines link essentially the size of the tumour ($R_0$) to the amplitude
of the RT reaction ($k$). The next two ones relate in a non-trivial
way, the growth speed ($v$) to the RT effect ($k,\tau)$.

\begin{figure}[!ht]
  \centering
  \includegraphics[width=0.7\textwidth]{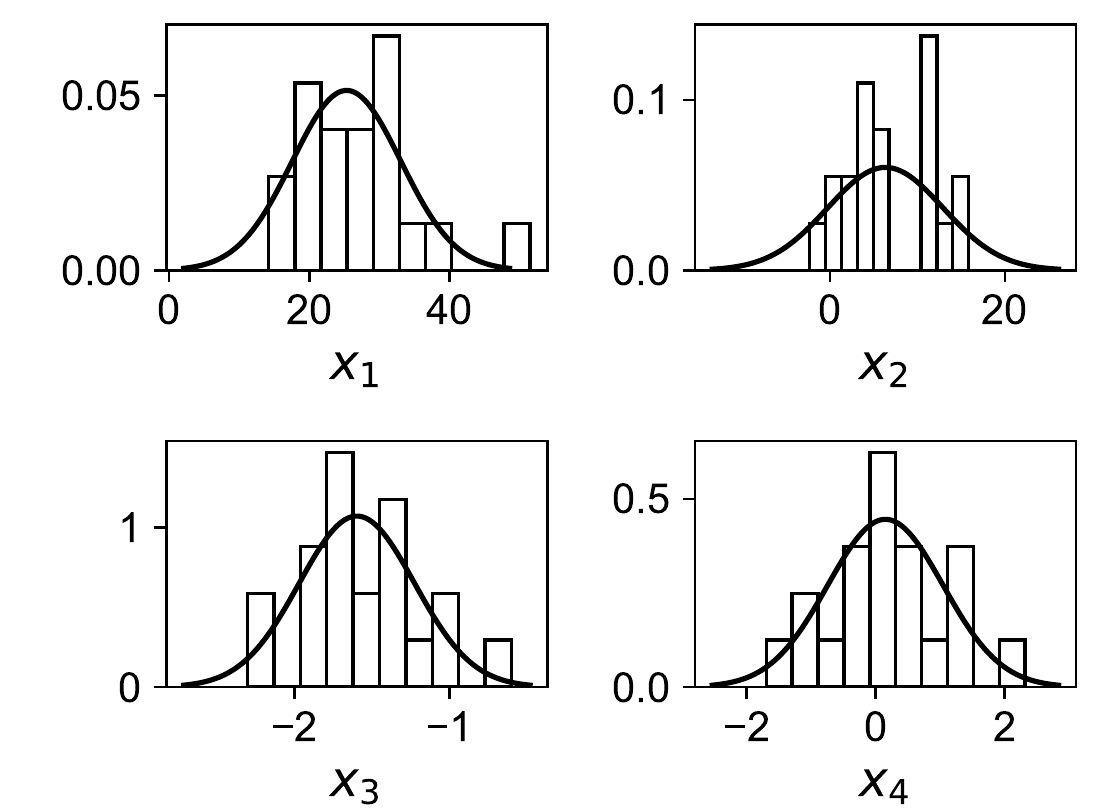}
  \caption{\textbf{Histograms of the transformed decorrelated variables}
    The new variables are linear combinations of the fitted (\params)
    parameters as 
    described in the text. They are normalized to unit area and the
    result of the Gaussian fit is shown in black.}
  \label{fig:xis}
\end{figure}

We now consider the distribution of these new $\{x_{i=1,\cdots,4}\}$ variables that
which, we recall, are mutually uncorrelated by construction. Their histograms
are shown on \Fig{xis}.

The nice feature now is that, unlike the original variables (\Fig{params}), the distributions are now approximately Gaussian
\footnote{although it is a slightly questionable assertion for $x_2$, the standard
  deviation of the fit is large enough to capture reasonably all the
  points.}.
For each variable we fit the mean ($\mu_i$) and standard-deviation $\sigma_i$.

We can now build a term that contains
the extra-information about the correlations among the variables in the form

\begin{align}
\label{eq:chi2cons}
  \chi^2_{cons}(\params)=\sum_{i=1}^4
  \left[\dfrac{(x_i(\params)-\mu_i)}{\sigma_i}\right]^2,
\end{align}
where the $x_i$'s are computed according to \refeq{prod}, and
($\mu_i,\sigma_i$) are the parameters of the Gaussian fits shown on \Fig{xis}.

We then add this term to the original $\chi^2$ function (\refeq{chi2})
\begin{align}
\label{eq:chi2tot}
  \chi^2_{TOT}(\params)=\chi^2(\params)+\chi^2_{cons}(\params)
\end{align}
and perform the minimization. The constraint acts as a
Bayesian prior, i.e. it includes all the a priori information we have 
between the parameters. It has no sense to use it on the previous fits
(\Fig{bf}) since it was derived from them.
But we have checked that the best-fits obtained using $\chi^2_{TOT}$
are \textit{exactly} the same as the ones with only $\chi^2$,
meaning that we are not over-constraining the parameters with the $\chi^2_{cons}$ term.

So why add such a term? Suppose we have few data, for
instance 2 measurements before RT and one after, then we have only 3 points
to determine 4 parameters. Using \refeq{chi2cons} we introduce
some extra equations and the problem becomes at least technically solvable.

In the following we focus on the regrowth time, which according
to our model (\refeq{model}) is
\begin{align}
\label{eq:tmin}
  \tmin=\tau\ln\left(\dfrac{k/\tau}{v}\right).
\end{align}
It depends mostly on $\tau$ and logarithmically on the relative
speed between the shrinkage due to RT ($k/\tau=v_d$) and the intrinsic tumour growth ($v$).
The $\chi^2_{TOT}(\params)$ minimization leads to the (\paramshat) estimates
and we use those values in \refeq{tmin} to estimate the regrowth time.

\section{Results}
\label{sec:pred}

\subsection{Data validation}
\label{sec:validation}
We first validate the procedure on our dataset by assessing the
performances of our predictions with a \textit{single} point after RT.

\begin{figure}[!ht]
  \centering
  \includegraphics[width=\textwidth]{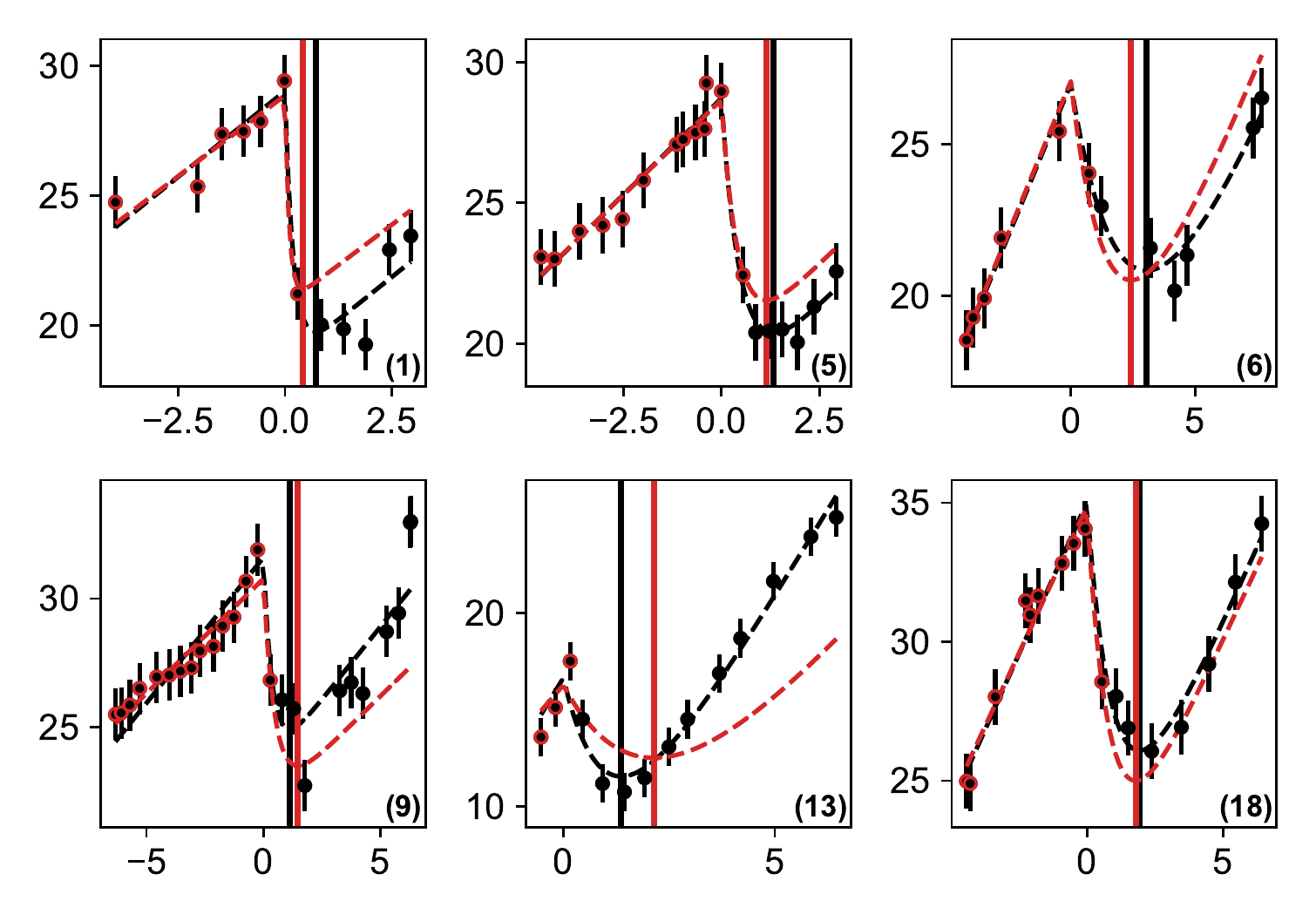}
  \caption{\textbf{Predictions for the regrowth time on real data with the
    first  point after RT}. The black dashed curve shows the best fit model using all
    the data points and the vertical black line shows its minimum
    (same as \Fig{bf}) . We then only
    consider the red circled points consisting of all the points before RT and the first
    one just after, and perform the constrained minimization described in
    the text. The result for the model is the dashed red curve with
    the estimated regrowth time shown as the vertical red line.}
  \label{fig:pred_data}
\end{figure}

Among our patients, we choose 6 follow-ups, with at least two points
before RT and enough subsequent points for the minimum of the fit to
be robust (see \Fig{bf}).
We then take the points before RT and the first one just after it, and 
perform the constrained minimization (\refeq{chi2tot}).
We obtain an estimate of \tmin and compare it to the one from the
full fit.
In order to avoid mixing the training and test samples, for each
patient we rebuild the constraint on the 
lines of section \ref{sec:cons}, removing each time the patient's data from the datasets.
The results are shown in \Fig{pred_data}.

The \tmin predictions for most of the patients are quite precise; they lie within
a few months of the value determined with all data. For patient
(13) it is slightly larger (10 months). This is 
an interesting case, since the point after RT is \textit{above} the
one before. This can be due to statistical fluctuations or to the fact that RT
produces sometimes an oedema that can be misidentified as the tumour radius.
However even in this case, we obtain a reasonable estimate.
This shows that, at the date of the first MRI after RT, we could
have guessed in most cases efficiently the regrowth time of the
tumour and plan more efficiently the dates of the next MRIs.

\subsection{Predictions}
\label{pred}

We now evaluate on virtual patients a strategy 
to estimate as soon as possible the tumour regrowth time. To this aim, we
must first fix the times of the MRI measurements which are constrained in the following way.

\begin{enumerate}
\item Although we showed results with many points before RT
(\Fig{pred_data}), today's clinical paradigm is to reduce the tumour
as soon as possible. We thus consider the case
where the radiotherapy sessions are planned immediately after the first
MRI within typically 6 months.
\item Since this is a central point, a second MRI should be performed
  around the RT date.
\item Fast-responders reacting within a few months, we propose
  to perform an MRI measurement 3 months after RT.
\end{enumerate}

\begin{figure}[!ht]
  \centering
  \includegraphics[width=.7\textwidth]{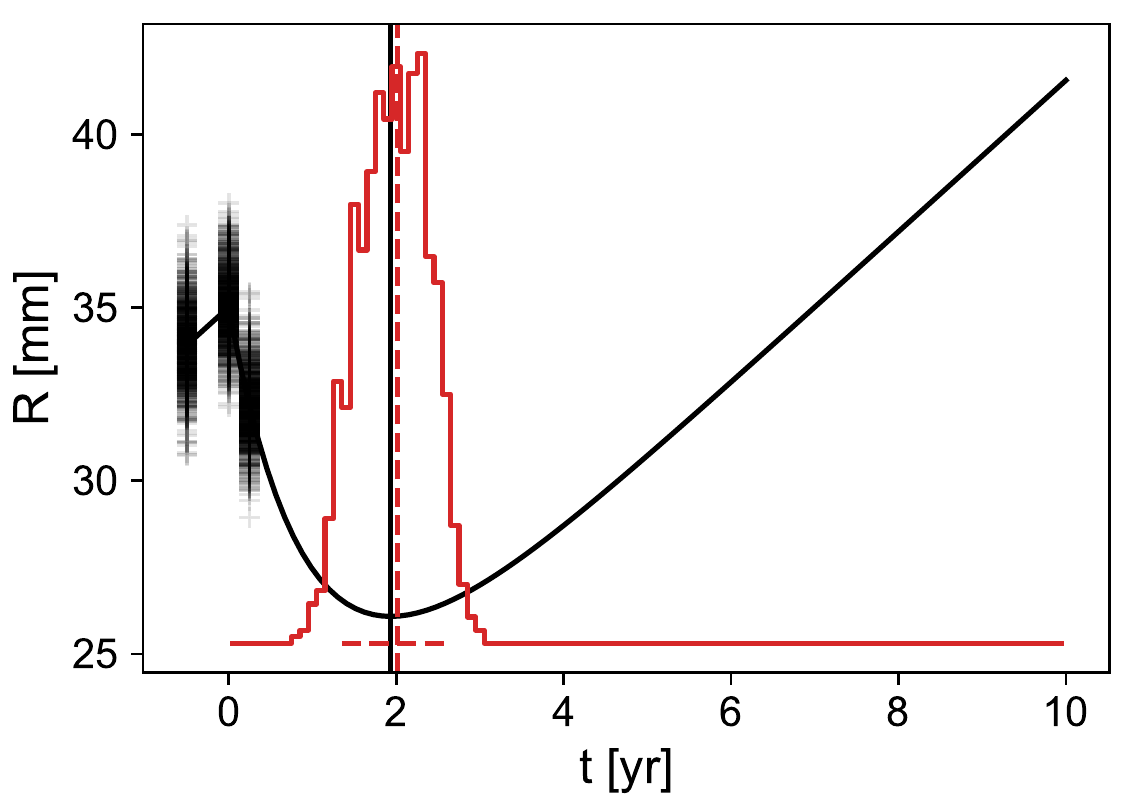}
  \caption{\textbf{Characterization of the
    regrowth time estimation with a Monte-Carlo method.} The black
  curve represents a model (which is here the bestfit of patient (18)) with its
    minimum shown as the vertical black line. One draws some
    Gaussian noise of $\sigma=1$ mm at the measurement times
    $\tmes=[-6,0,3]$ months, and performs the \tmin estimation described
    in the text. This is repeated 1000 times which allows to construct
    the red histogram of all the \tmin estimates. The red vertical dashed line
    shows its mean value and the horizontal one the [0.05,0.95]
    percentile region.
  }
  \label{fig:mc_example}
\end{figure}

We will then consider the cases where the measurement times are
located at $\tmes=[-6,0,+3]$ months and test if we can still 
make some predictions for the regrowth time. This is a very challenging situation
since we only have 3 nearby points with important relative errors.
To assess statistically the performances of the prediction, we adopt a Monte-Carlo approach.
For a given set of ``true'' parameters (\params), we first compute the 
tumour radius at $\tmes$. We then add to each point 
a random Gaussian noise with a $\sigma=1$ mm standard deviation, 
and from these virtual measurements, estimate the regrowth time. 
We repeat the procedure 1000 times and
consider the mean of the predictions and the 95\% confidence-level
interval (obtained from the $[0.05,0.95]$ percentiles) that we compare to the true \tmin value.
This procedure is illustrated in \Fig{mc_example}.

\begin{figure}[!ht]
  \centering
  \includegraphics[width=.7\textwidth]{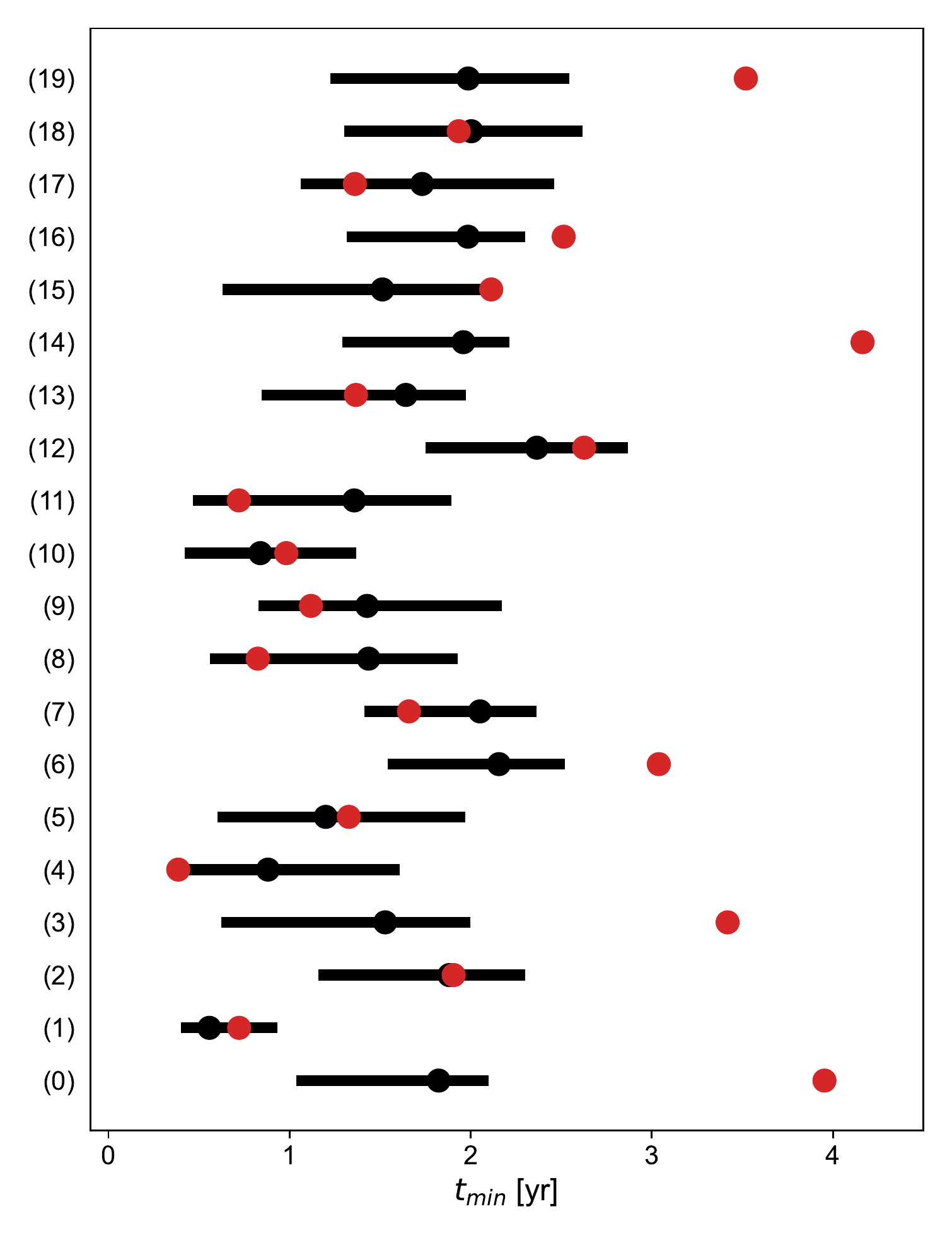}
  \caption{\textbf{Performances of the
    regrowth time estimates with 3 measurements at $\tmes=[-6,0,3]$
    months for a set of true parameters corresponding to the bestfits
    of our 20 patients.}
  Black points represent  the mean of the estimates and
    the bars the 95\% confidence-level interval. The red point is the
    true value associated to each best-fit for the patients labelled on
    the vertical axis  (corresponding to the dashed lines in \Fig{bf}).
    Note that only the best-fit parameters of each patient are being used here.}
  \label{fig:pred3mois}
\end{figure}

We use our 20 best-fits as
a representative set of ``true models''.
We perform the Monte-Carlo study described previously for each set
of parameters and compare the mean and 95\% confidence-level interval
of our estimated regrowth times to the true value on \Fig{pred3mois}.

First, we notice that 15 predictions out of 20 (75\%) are good, the
mean value being typically within 6 months of the true. In these
cases, the guess follows roughly the true values which confirms that
the method is not only driven by the constraint (which would lead
always to the same interval) but also incorporates the information of
the 3 measurements. 
Fast-responders (patients (1),(4),(8) and (11)) are correctly predicted and tend to
lead to predictions under 1 year which could be the threshold to plan
a next MRI rapidly (possibly 3 months later).

There are also 5 outliers out of 20 (25\%)
corresponding to the cases where the true regrowth times are the
largest ($3-4$ years), i.e. to the \textit{slowest} responders. 
A point at 3 months for them is much too soon to infer \textit{any} information
about the curvature, so that the prediction is only driven by the
constraint and goes to its mean value of about 2 years.
More precisely, by Taylor-expanding our model near $t=0^+$
\begin{align}
\label{eq:dl}
  R(t)=R_0+(v-v_d)t + \epsilon t^2/2 +\bigO{t^3}
\end{align}
where  $v_d\equiv k/\tau$ is the speed of the collapse and the
curvature term is $\epsilon=v_d/\tau$.
For slow-responders, there is almost no curvature at 3 months,
$\epsilon\to0$ and $\tau=v_d/\epsilon$ diverges leading to a very broad (and even
sometimes bi-modal) \tmin distribution. In this case, the prediction
is only driven by the constraint. 

\begin{figure}[!ht]
  \centering
  \includegraphics[width=0.7\textwidth]{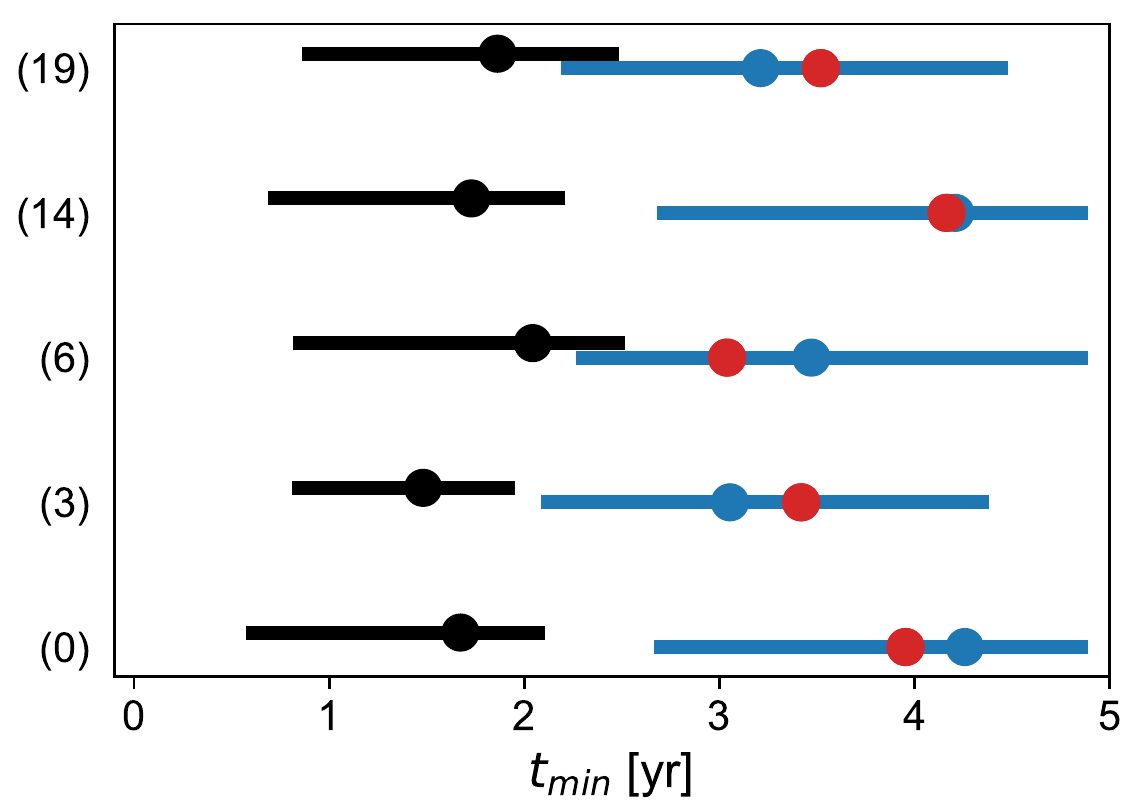}
  \caption{\textbf{Monte-Carlo estimates of the regrowth time
      with an extra point at 1 year for the 5 outliers of
      \Fig{pred3mois} ($\tmin>3$ years).} 
    Black bars corresponds to the 95\% confidence-level intervals
    obtained from the standard constraint (\refeq{chi2cons}) and the blue
    ones with the loose constraint described in the text. The black/blue
    points corresponds to the mean values and the red point is the true
    value of each model.
}

  \label{fig:pred1an_new}
\end{figure}

Although pessimistic for the patient, the predicted value is still \textit{large} (around 2 years, see
\Fig{pred3mois}). Thus we can safely plan a next MRI 1 year after RT.
We consider the case  where the times for the radial measurement are at
$\tmes=[-6,0,3,12]$ months and perform the prediction again. The
result is shown in black in \Fig{pred1an_new}.

\begin{figure}[!ht]
  \centering
  \includegraphics[width=0.7\textwidth]{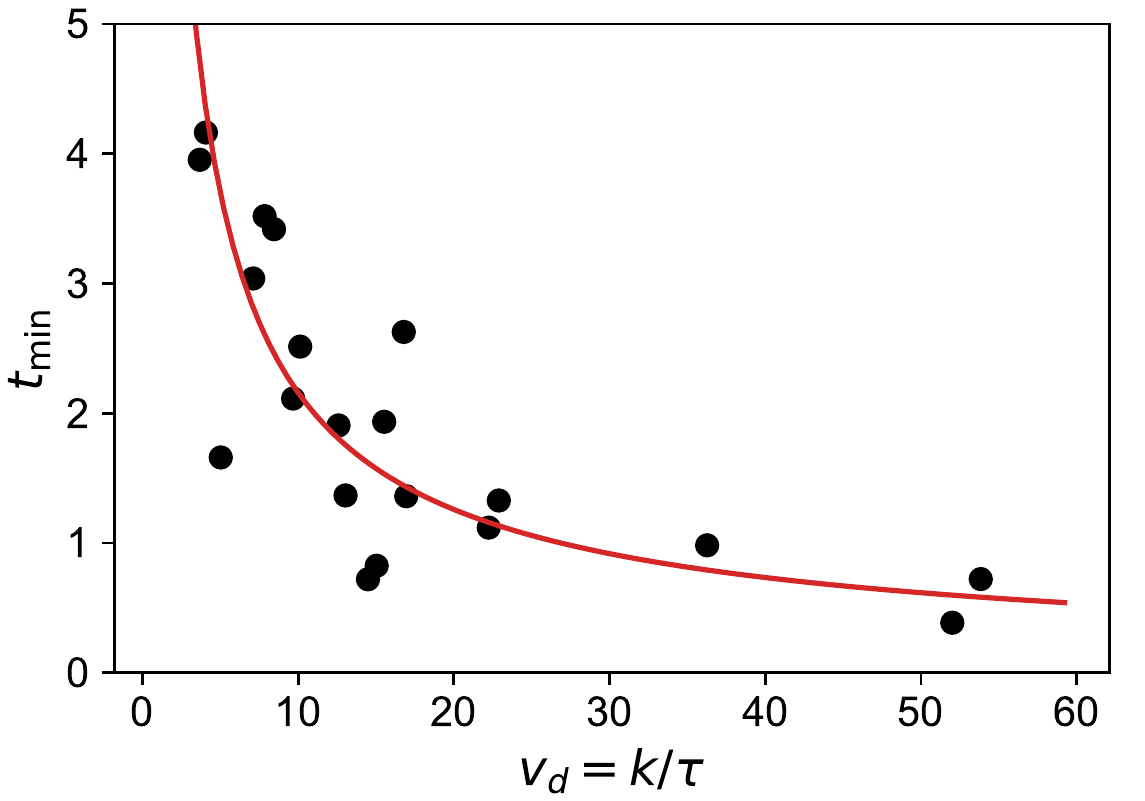}
  \caption{\textbf{Correlation between $v_d$ and \tmin measured on our set of
    20 patients.} The red line shows the power-law fit.}
  \label{fig:corrvdtmin}
\end{figure}

Unfortunately, the constraint \refeq{chi2cons} is
still pulling \tmin to too low values. We need to switch to a looser constraint.
As is clear from \refeq{dl}, the linear term, that is the best
constrained, is related to the slopes
measured before ($v$) and after ($v_d$) RT. In the absence of good
knowledge of the curvature, we may try to relate these slopes
to the regrowth time.
Indeed, on our dataset, we observe a strong correlation
between $v_d$ and $\tmin$ (\Fig{corrvdtmin}) that we fit to a power-law 
\begin{align}
\label{eq:vd2tmin}
  \tmin=13/v_d^{0.78}.
\end{align}

The origin of this heuristic constraint remains to be understood, but we can use it
to build a new estimator for \tmin: 
we fit on the data only the linear terms in \refeq{dl} in order to get $v_d$
which we transform according to \refeq{vd2tmin}. Since this method uses a
single correlation, we call it the \textit{loose constraint}.
We show the result of applying this procedure on the outliers in blue
on \Fig{pred1an_new}. The distributions are much better centred on the true values. 
One may ask why not always use this constraint.
As clear in the figure, the uncertainty is larger with the
loose-constraint method. This is the classical bias/variance trade-off of any estimator. 
Although it gives indeed less biased results for outliers, the method
would miss fast-responders at 3 months, since the slopes determination
is then extremely noisy. On the contrary, with a point at 1 year, there is enough lever-arm
to determine the slope quite precisely and take advantage from the
correlation to let the data ``speak for themselves''.

We point out that the loose-constraint method is very simple and may be used by
any clinician without even a computer. First measure the slope
before RT to obtain $v$, then the slope after RT ($v-v_d$) to obtain $v_d$,
and finally use \refeq{vd2tmin} to predict the regrowth-time.

\section{Discussion}

We have proposed a new simple model to describe the evolution of
diffuse low-grade gliomas before and after radiotherapy. It is
analytical and describes in a satisfactory way the follow-ups of 20 patients
with measured tumour radii before and after RT.
This model has 4 free parameters, 2 before RT and 2 after, that vary for each patient.
From the study of the correlation between all the parameters we 
proposed a way to include a prior information to any follow-up, which
allows to perform predictions for the regrowth-time of the tumour
rapidly after RT.
From the data we had at our disposal, 
we showed that including this information
allows to predict the regrowth time of the tumour at the very first MRI
measurement after RT typically within 6 months.
Using virtual patients, we have shown that is is possible to
predict reasonably well the regrowth time with only one point 6 months before RT, one
around RT and one 3 months after, in 75\% of the cases. The remaining
25\% for which our prediction is \textit{pessimistic}, have all large regrowth-time ($\simeq$ 4 years) and may 
draw benefit from another measurement 1 year after RT, leading to more
correct estimates.

These results assume that our database is
representative of all LGG evolution and would profit from incorporating
more patients' data. Similar profiles are obtained for chemotherapy
treatments \cite{Mazzocco:2015,Bodnar:2019} 
and it would be interesting to redo the analysis in this case.

This work is based on a 4-parameters model which is a
simplified version of a biologically motivated model. This 
choice can be challenged; why not use some non-parametric method that are often efficient?
First, the low dataset (43 patients but in practice 20 with a
sufficient number of points to inform our model) precludes the
possibility of using  general purpose Machine Learning techniques 
like Deep Neural Networks, Random Forests, Boosted Decision Trees (as
described for instance in this recent review  \cite{math9222970}), as
well as Recurrent Networks dedicated to Time Series (e.g. \cite{ijerph19042417}). 
Second, we could think of using Gaussian Processes (GP) method (e.g. \cite{forecast3010013})
that can work on small samples  with some
optimized kernel. We have tried it, with a squared exponential kernel
and a white noise. However, by construction, outside the
data input region the naive ``vanilla'' model converges to a constant and cannot
describe the regrowth phase. 
To overcome this failure, one is forced to use a time dependent
function of the mean which is exactly the meaning of the 4-parameters
model developed in this article. 
This clarifies why modelling, especially based on physical arguments,
is superior to all purely statistical methods. This was the key to the
success of making predictions from a restricted dataset and with very few data points.

Here, we have varied the
patient's population and shown that the method has the potential
to make some predictions among various patients profile.
The problem is different for a personalised follow-up (which is the
practical clinical case) since the prediction  
depends on the details of the measurements (times and values). Using a Monte-Carlo
Markov Chain technique, one can obtain an individualised probability
distribution of the regrowth-time that can help clinicians adapt their
treatment and the dates of the next MRIs. We plan to provide such a
tool that will be publicly available online.

\nolinenumbers

%
%
%


\end{document}